\def\papertitle{Differentiable time--frequency scattering on GPU}
\def\paperauthorA{John Muradeli}
\def\paperauthorB{Cyrus Vahidi}
\def\paperauthorC{Changhong Wang}
\def\paperauthorD{Han Han}
\def\paperauthorE{Vincent Lostanlen}
\def\paperauthorF{Mathieu Lagrange}
\def\paperauthorG{George Fazekas}

\documentclass[twoside,a4paper]{article}
\usepackage{etoolbox}
\usepackage{dafx_22a}
\usepackage{amsmath,amssymb,amsfonts,amsthm, bm}
\usepackage{euscript}
\usepackage[T1]{fontenc}
\usepackage[utf8]{inputenc}
\usepackage{microtype}
\usepackage{nimbusserif}
\usepackage{ifpdf}
\usepackage[english]{babel}
\usepackage{caption}
\usepackage{subfig} % or can use subcaption package
\usepackage{color}
\usepackage{multirow}

\input glyphtounicode
\ninept

\newcounter{numauth}
\newcounter{listcnt}
\newcommand\authcnt[1]{\ifdefined#1 \stepcounter{numauth} \fi}

\newcommand\addauth[1]{
\ifdefined#1 
\stepcounter{listcnt}
\ifnum \value{listcnt}<\value{numauth}
\appto\authorslist{, #1}
\else
\appto\authorslist{~and~#1}
\fi
\fi}
\authcnt{\paperauthorB}
\authcnt{\paperauthorC}
\authcnt{\paperauthorD}
\authcnt{\paperauthorE}
\authcnt{\paperauthorF}
\authcnt{\paperauthorG}
\authcnt{\paperauthorH}
\authcnt{\paperauthorI}
\authcnt{\paperauthorJ}
\def\authorslist{\paperauthorA}
\addauth{\paperauthorB}
\addauth{\paperauthorC}
\addauth{\paperauthorD}
\addauth{\paperauthorE}
\addauth{\paperauthorF}
\addauth{\paperauthorG}
\addauth{\paperauthorH}
\addauth{\paperauthorI}
\addauth{\paperauthorJ}
\usepackage{times}
\newif\ifpdf
\ifx\pdfoutput\relax
\else
   \ifcase\pdfoutput
      \pdffalse
   \else
      \pdftrue
\fi

\ifpdf % compiling with pdflatex
  \usepackage[pdftex,
    pdftitle={\papertitle},
    pdfauthor={\authorslist},
    pdfsubject={Proceedings of the 25th International Conference on Digital Audio Effects (DAFx20in22)},
    colorlinks=false, % links are activated as color boxes instead of color text
    bookmarksnumbered, % use section numbers with bookmarks
    pdfstartview=XYZ, % start with zoom=100% instead of full screen; especially useful if working with a big screen :-
  ]{hyperref}
  \usepackage[pdftex]{graphicx}
\else % compiling with latex
  \usepackage[dvips]{epsfig,graphicx}
  \usepackage[dvips,
    pdftitle={\papertitle},
    pdfauthor={\authorslist},
    pdfsubject={Proceedings of the 25th International Conference on Digital Audio Effects (DAFx20in22)},
    colorlinks=false, % no color links
    bookmarksnumbered, % use section numbers with bookmarks
    pdfstartview=XYZ % start with zoom=100% instead of full screen
  ]{hyperref}
\fi
\usepackage[hypcap=true]{caption}
\title{\papertitle}

\affiliation{\paperauthorA$^*$, \paperauthorB$^\#$, \paperauthorC$^+$, \paperauthorD$^+$, \paperauthorE$^+$, \\ \paperauthorF$^+$, \paperauthorG$^\#$ \thanks{Cyrus Vahidi is a researcher at the UKRI CDT in AI and Music, supported jointly by the UKRI (grant number EP/S022694/1) and Music Tribe. This work was conducted while at LS2N, CNRS.
Changhong Wang is supported by an Atlanstic2020 project on Trainable Acoustic Sensors (TrAcS).}
}
{
\begin{tabular}{ccc}
  & $^+$LS2N, CNRS, Nantes Université & $^\#$Center for Digital Music \\
  & École Centrale Nantes, France &  Queen Mary University of London, UK  \\
$^*$\tt john.muradeli@gmail.com & \tt  firstname.lastname@ls2n.fr & \tt  f.lastname@qmul.ac.uk
\end{tabular}
}

\usepackage{amsmath}

\DeclareMathOperator*{\argmin}{arg\,min}
\usepackage{siunitx}
\usepackage{url}
\begin{document}
% more pdf-tex settings:
\ifpdf % used graphic file format for pdflatex
  \DeclareGraphicsExtensions{.png,.jpg,.pdf}
\else  % used graphic file format for latex
  \DeclareGraphicsExtensions{.eps}
\fi

%\makeatletter
%\pdfbookmark[0]{\@pdftitle}{title}
%\makeatother

\maketitle

\begin{abstract}
Joint time--frequency scattering (JTFS) is a convolutional operator in the time--frequency domain which extracts spectrotemporal modulations at various rates and scales.
It offers an idealized model of spectrotemporal receptive fields (STRF) in the primary auditory cortex, and thus may serve as a biological plausible surrogate for human perceptual judgments at the scale of isolated audio events.
Yet, prior implementations of JTFS and STRF have remained outside of the standard toolkit of perceptual similarity measures and evaluation methods for audio generation.
We trace this issue down to three limitations: differentiability, speed, and flexibility.
In this paper, we present an implementation of time--frequency scattering in Python.
Unlike prior implementations, ours accommodates NumPy, PyTorch, and TensorFlow as backends and is thus portable on both CPU and GPU.
We demonstrate the usefulness of JTFS via three applications: unsupervised manifold learning of spectrotemporal modulations, supervised classification of musical instruments, and texture resynthesis of bioacoustic sounds.
\end{abstract}

\section{Introduction}
\label{sec:intro}
Human listening plays a central role in the development and evaluation of digital audio effects (DAFx)  \cite{pressnitzer2000acoustics}.
Yet, listening tests are costly and time-consuming as they typically rely on expert participants.
For this reason, recent publications have proposed to mimic the behavioral response of the average listener by means of a computational surrogate \cite{lostanlen2021time,thoret2021learning}.
In particular, experimental findings in auditory neurophysiology suggest that our primary cortex responds selectively to spectrotemporal modulations at various rates and scales \cite{depireux2001spectro}.
Each of these responses may be simulated by an idealized model known as spectrotemporal receptive field (STRF).
Hence, the space of STRF coefficients appears as a natural candidate for comparing two sounds out of context; and indeed, studies in music psychology have confirmed that Euclidean distances in STRF space approximate timbre dissimilarity judgments between isolated musical notes \cite{patil2012music}.

%\cite{martinez2019general}
However, despite its potential for developing perceptually informed audio synthesis, the STRF has received limited adoption within the DAFx community as well as music information retrieval (MIR) and machine learning for signal processing (MLSP).
Indeed, we notice three shortcomings in the NSL Auditory--Cortical Toolbox \cite{chi2005multiresolution}, which we consider to be the reference implementation of STRF \footnote{\url{http://nsl.isr.umd.edu/downloads.html}}.
First, it lacks scalability: the code is written in MATLAB, does not accommodate parallel computing, and is not portable onto GPU hardware.
Second, it lacks flexibility: the toolbox assumes that the audio input has a sample rate of \SI{16}{\kilo\hertz} and  subsamples all subbands in the constant-$Q$ filterbank to \SI{125}{\hertz}, even though the critical sample rate should depend on center frequency so as to minimize memory usage while avoiding aliasing artifacts.
Thirdly, it lacks differentiability: although the authors do provide a modified Griffin-Lim algorithm for reconstructing an audio signal from its STRF coefficients, MATLAB's NSL and related toolboxes do not perform reverse-mode automatic differentiation, unlike PyTorch or TensorFlow.
The same three issues are found again in the toolbox ``strf-like-model'' of \cite{thoret2021learning}, which is a Python port of the NSL Auditory--Cortical Toolbox using NumPy as its backend \footnote{\url{https://github.com/EtienneTho/strf-like-model}}.

\begin{figure*}[ht]
    \centering
    \includegraphics[width=.24\linewidth]{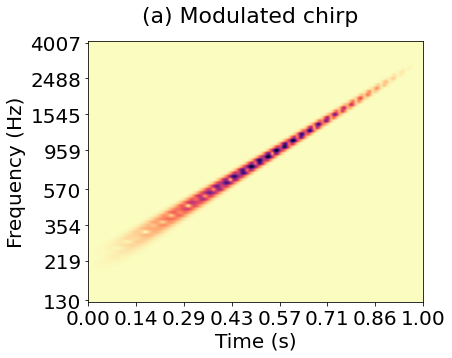} \
    \includegraphics[width=.24\linewidth]{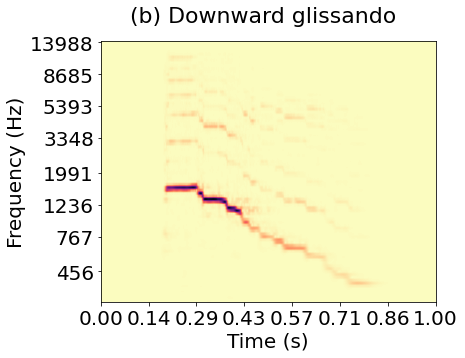} \
    \includegraphics[width=.24\linewidth]{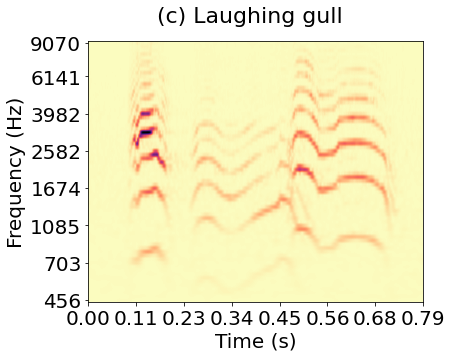}\
    \includegraphics[width=.24\linewidth]{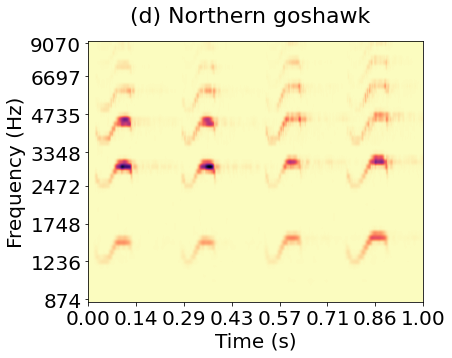}\\
    \includegraphics[width=.24\linewidth]{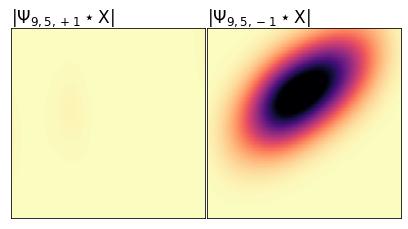} \
    \includegraphics[width=.24\linewidth]{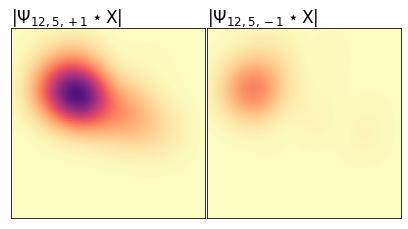} \
    \includegraphics[width=.24\linewidth]{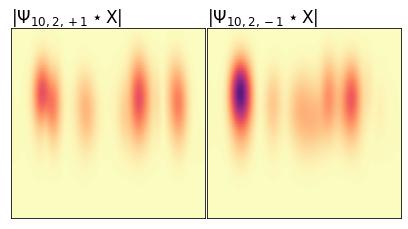} \
    \includegraphics[width=.24\linewidth]{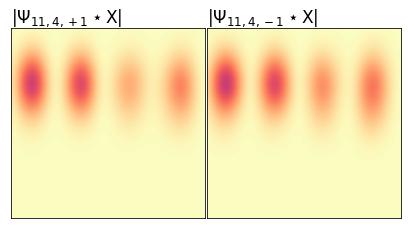}
    \caption{Scalograms (top) and scale--rate visualizations (bottom) from the joint time--frequency scattering transforms of exemplary audio signals: (a) amplitude-modulated chirp signal; (b) downward glissando musical instrument playing technique; (c) and (d) are the sounds of two types of birds, the laughing gull and the Northern goshawk. Each scale--rate visualization shows the response of a second-order 2-D wavelet $\psi$ of temporal rate $\alpha$, frequential scale $\beta$ and orientation $\theta =\pm 1$ when convolved with the scalogram $X$. See Section \ref{sec:jtfs} for details on joint time--frequency scattering.}
    \label{fig:JTFS_example_visual}
\end{figure*}

In this paper, we present a Python implementation of ``Joint Time-Frequency Scattering'' (JTFS) \cite{Anden2019joint}, that is, a fast and numerically accurate discretization of STRF.
While there have been implementations of JTFS in MATLAB since 2014\footnote{\url{https://www.di.ens.fr/data/software/scatnet} }~\footnote{\url{https://github.com/lostanlen/scattering.m}}, ours is the first to support GPU computing and automatic differentiation.
To accomplish this, we have extended Kymatio\footnote{\url{https://www.kymat.io/}}, a library for wavelet-based processing in Python released in 2019 \cite{andreux2020kymatio}, with code that now constitutes a permanent branch called dafx2022-jtfs\footnote{Time-frequency scattering: \url{https://github.com/OverLordGoldDragon/wavespin/tree/dafx2022-jtfs}}, in WaveSpin, a library currently under development. 
We show the potential of the implementation to different research topics with three examples: unsupervised manifold learning of spectrotemporal modulations, supervised classification of musical instruments, and texture resynthesis of bioacoustic sounds. Beyond the demonstrated use cases, differentiable implementations of scattering have potential to enable parametric scattering filterbanks \cite{gauthier2022parametric} and audio synthesis loss functions.
Our supervised classifier is the first instance of ``hybrid representation learning'' \cite{oyallon2018scattering} which interfaces JTFS with a 2-D deep convolutional network (convnet). Furthermore, we outline an activation function for scattering-based neural networks, under the name of mean-based logarithm ($\mu$--$\log$). We demonstrate state-of-the-art musical instrument classification results in the setting of limited annotated training data. For the sake of reproducibility, we provide open-source code, with experiments reproduced in a repository named JTFS-GPU\footnote{Experiments repository: \url{ https://github.com/cyrusvahidi/jtfs-gpu}}, alongside supplementary material\footnote{Companion website: \url{https://cyrusvahidi.github.io/jtfs-gpu}}.

% Furthermore, we introduce a new activation function for scattering-based neural networks, under the name of \emph{Adaptive Logarithm} (AdaLog).
% We observe that AdaLog outperforms conventional (constant-based and mean-based) logarithms in a supervised task of musical instrument classification.

\section{Time--Frequency Scattering} \label{sec:jtfs}
Proposed in \cite{Anden2019joint}, the \emph{joint time--frequency scattering} (JTFS) transform captures the spectrotemporal modulations of a signal at various rates and scales. 
This is achieved by decomposing the signal with joint wavelet convolutions, nonlinearities, and pooling operations. 

Let ${\bm \psi}(t)$ and ${\bm \psi}(\lambda)$ denote the basis function (``mother wavelet'') for the decomposition along the time, $t$, and the log-frequency axis, $\lambda$, respectively.
Both wavelets are complex analytic, of which the Fourier transform is null for negative frequencies, i.e. $\hat{\bm \psi}(\omega)=0$ for $\omega<0$.
Our implementation uses the Morlet wavelet, i.e. a complex sinusoid modulated by a Gaussian envelope, due to its quasi-optimality in terms of Heisenberg time--frequency uncertainty.
${{\bm \psi}}_{\lambda}(t)$ is the temporal wavelet filterbank dilated from ${\bm \psi}(t)$.
Convolving a waveform ${\bm x}(t)$ with each wavelet in ${\bm \psi}_{\lambda}(t)$ and applying pointwise complex modulus yields the scalogram $\mathbf{X}(t,{\rm \lambda}) = \big| {\bm x} \ast {{\bm \psi}}_{\lambda} \big|(t)$,
which is a two-dimensional (2-D) time--frequency image, as shown by the examples in Fig.~\ref{fig:JTFS_example_visual} (top).

To extract the spectrotemporal modulations of an STRF centered at $(t,\lambda)$, we decompose the scalogram with a joint time--frequency wavelet $\mathbf{\Psi}_{\alpha,\beta,\theta}(t,\lambda)$. $\alpha$ and $\beta$ are the temporal rate and frequential scale, respectively. 
$\theta=\pm 1$ is the orientation of the STRF, with $\theta=-1$ denoting a positive slope while $\theta=+1$ a negative one.
${\bm \psi}_{\alpha}(t)$ and ${\bm \psi}_{\beta}(\lambda)$ are the temporal and the frequential wavelet filterbank dilated from their mother wavelets, ${\bm \psi}(t)$ and ${\bm \psi}(\lambda)$, respectively:
\begin{eqnarray}
    {\bm \psi}_{\alpha}(t)&=&2^{\alpha}{\bm \psi}(2^{\alpha}t) \hspace{0.2in} \mbox{and}\nonumber \\
    {\bm \psi}_{\beta,\theta}(\lambda)&=&2^{\beta}{\bm \psi}(\theta2^{\beta}\lambda).
\end{eqnarray}
The joint time--frequency wavelet $\mathbf{\Psi}_{\alpha,\beta,\theta}(t,\lambda)$ is the outer product between the temporal wavelet ${\bm \psi}_{\alpha}(t)$ and the frequential wavelet ${\bm \psi}_{\beta,\theta}(\lambda)$:
\begin{equation}\label{eq:jointWavelets}
    \mathbf{\Psi}_{\alpha,\beta,\theta}(t,\lambda) =
    {\bm \psi}_{\alpha}(t)
    {\bm \psi}_{\beta,\theta}(\lambda).
\end{equation}

We then convolve the scalogram with the 2-D joint wavelet filterbank $\mathbf{\Psi}_{\alpha,\beta,\theta}(t,\lambda)$, apply a complex modulus, and average it by a 2-D lowpass filter ${\bm \Phi}_{T,F}(t,\lambda)$.
Following~\cite{Anden2019joint}, we define the joint time--frequency scattering of $\mathbf{X}(t, \lambda)$ as:
\begin{equation}\label{eq:jTFSTDefine}
    \mathbf{S}_{2}^{\rm JTFS}{\bm x}(t, \lambda, \alpha, \beta, \theta) = \Big( \big| \mathbf{X} \overset{t,\lambda}{\ast} \mathbf{\Psi}_{\alpha,\beta,\theta} \big| \overset{t,\lambda}{\ast} \mathbf{\Phi}_{T,F} \Big)(t,\lambda).
\end{equation}
The symbol $\overset{t,\lambda}{\ast}$ denotes a 2-D convolution over both the time variable $t$ and the log-frequency variable $\lambda$.
Hence, for each location around $(t, \lambda)$ in the time--frequency domain, we obtain a 3-D tensor indexed by $(\alpha, \beta, \theta)$, capturing spectrotemporal modulation information.
$\mathbf{S}_{2}^{\rm JTFS}{\bm x}$ in Eq.~(\ref{eq:jTFSTDefine}) has invariance properties to time-shifts, time-warps, and frequency transpositions, for a receptive field restricted by the time scale $T$ and frequency interval $F$.
In certain cases, we may omit frequential averaging in order to preserve equivariance to frequency transposition.
This results in a variant of Eq. (\ref{eq:jTFSTDefine}):
\begin{equation}\label{eq:jTFSTDefine-t}
    \mathbf{S}_{2}^{\rm JTFS}{\bm x}(t, \lambda, \alpha, \beta, \theta) = \Big( \big| \mathbf{X} \overset{t,\lambda}{\ast} \mathbf{\Psi}_{\alpha,\beta,\theta} \big| \overset{t}{\ast} \mathbf{\Phi}_{T} \Big)(t,\lambda).
\end{equation}

Fig. \ref{fig:JTFS_example_visual} shows the scalogram (top) and the scale--rate visualizations of JTFS (bottom) for four audio examples. (a) is a synthesized amplitude-modulated chirp signal of constant chirp rate, see Section \ref{sec:dataset} for details. (b) is a glissando playing technique performed on the Chinese bamboo flute \emph{(dizi)}; (c) and (d) are the vocalisations of two species of birds: the laughing gull and the Northern goshawk \emph{(Accipiter gentilis)}. 
The first two examples are isolated events with upward and downward frequency change, respectively, while the remaining two are real-world acoustic events with temporal variations in chirp rate and directionality.

In Fig. \ref{fig:JTFS_example_visual}, we visualize JTFS coefficients before averaging, i.e. $|\mathbf{X} \ast \mathbf{\Psi}_{\alpha, \beta, \theta}|$, with one scale--rate combination for each of the audio examples.
All visualizations cover the complete time $t$ and log-frequency $\lambda$ axes, and both directions of $\theta$.
For instance, Fig. \ref{fig:JTFS_example_visual} (a) bottom displays the JTFS obtained by convolving $\mathbf{X}(t,\lambda)$ with $\mathbf{\Psi}_{9,5,-1}$ and $\mathbf{\Psi}_{9,5,1}$ respectively, taking complex modulus and lowpass filtering. $9$ and $5$ denote the temporal rate and frequential scale indices, respectively.
The scale--rate visualizations of (b), (c), and (d) are obtained in the same way.
As can be seen in the first two isolated examples (a) and (b), the direction of the spectrotemporal patterns are clearly captured: the JTFS energy concentrates on $\theta=-1$ for upward frequency changes in (a), while for the downward frequency variations in (b), $\theta=+1$ dominates the spectrotemporal modulations. The main directions of frequency modulations of each isolated event in examples (c) and (d) are also captured. 
We explore these two examples further in Section \ref{sec:textureresynth} in a task of audio texture resynthesis via JTFS coefficients.

\section{Similarity retrieval}
\subsection{Motivation}
In this section, we compare the abilities of audio representations to serve as a similarity measure between audio signals with real-world factors of variability.
We design a sound synthesizer to generate a dataset of \emph{amplitude modulated chirp} (AM/FM) signals, that is controlled by three parameters: carrier frequency ($f_c$, in Hz), amplitude modulation frequency ($f_m$, in Hz) and chirp rate ($\gamma$, in octaves/second).
Such amplitude and frequency modulations are typically found within musical instrument playing techniques \cite{wang2021joint}. We visualize similarity of the synthesized signals on a manifold embedding and assess recovery of the synthesizer parameters under several audio representations: Mel-frequency cepstral coefficients (MFCCs), time scattering (Scattering1D), time--frequency scattering (JTFS), spectrotemporal receptive fields (STRFs) and OpenL3 embeddings.
MFCCs result from computing a log-mel spectrogram (logmelspec) followed by a discrete cosine transform (DCT). The cortically-inspired spectrotemporal receptive field (STRF) transformation serves as a representation of spectrotemporal modulations \cite{thoret2021learning}.
OpenL3 embedding is a deep feature representation that results from training L3-Net for audiovisual correspondence \cite{cramer2019look}. 

\subsection{Synthetic Dataset of Modulated Chirps} \label{sec:dataset}
We define a generator $\boldsymbol{g}$ of exponential ``chirps'' with three factors of variability: a carrier frequency $f_c$, an amplitude modulation (AM) frequency $f_m$, and a frequency modulation (FM) rate $\gamma$.
Denoting by $\boldsymbol{\theta}$ the triplet $(f_c, f_m, \gamma)$, we have for every $\boldsymbol{\theta}$:
\begin{equation} \label{eqn:synth}
    \boldsymbol{g_\theta}: t\longmapsto
    \boldsymbol{\phi}_{w}(\gamma t)
    \sin(2\pi f_\mathrm{m} t)
    \sin\left(
    \dfrac{2\pi f_{\mathrm{c}}}{\gamma \log 2} 2^{\gamma t}
    \right),
\end{equation}
where $\boldsymbol{\phi}_{w}$ is a Gaussian window of characteristic width equal to $w$.
The AM/FM signal $\boldsymbol{g_\theta}$ has an instantaneous frequency of $f_c 2^{\gamma t}$ and an essential duration of $w/\gamma$.
Thus, it covers a bandwidth $w$, independently from $\boldsymbol{\theta}$.
We set $w = 2$ octaves in the following.

We highlight a physical correspondence of the synthesizer parameters to the wavelet variables in Section \ref{sec:jtfs}. $f_c$ corresponds to the log-frequency $\lambda$ of first-order scattering wavelets. Amplitude modulation frequency $f_{m}$ can be adequately described by second-order temporal wavelet rate $\alpha$. The chirp rate $\gamma$ accounts for a relationship between frequential wavelets of scale $\beta$ and rate $\alpha$, i.e. $\beta = \frac{\alpha}{\gamma}$.

\begin{figure}[h!]
    \centering
    \includegraphics[width=.8\columnwidth]{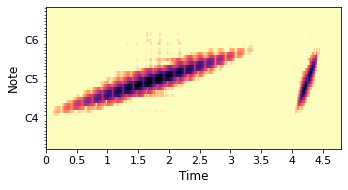}
    \caption{constant-$Q$ transform of two AM/FM signals with chirp rate $\gamma = 0.5$ (left) and $\gamma = 4$ (right) octaves/second, respectively. Both cases have a carrier frequency $f_c$ of \SI{512}{Hz}, a modulation frequency $f_m$ of \SI{3}{\hertz}, and a bandwidth equal to two octaves.}
    \label{fig:synth_chirps}
\end{figure}

We apply Eq. (\ref{eqn:synth}) for $16$ values of $f_c$, $f_m$, and $\gamma$, arranged in a geometric progression; hence yielding a dataset of $16^3=4096$ audio signals in total.
We vary $f_c$ between \SI{512}{\hertz} to \SI{1024}{\hertz}; $f_m$, between and \SI{4}{\hertz} to \SI{32}{\hertz}; and $\gamma$, between 0.5 and 4 octaves/second respectively.
Fig. \ref{fig:synth_chirps} illustrates the constant-$Q$ transform of two chirp signals of \SI{512}{\hertz} fundamental frequency, with chirp rates $\gamma = 0.5$ and $\gamma = 4$ octaves per second and equal bandwidth 2 octaves\footnote{See companion website: \url{https://cyrusvahidi.github.io/jtfs-gpu}}.

\subsection{Manifold Learning and Visualization} \label{sec:isomap}
To visualize similarity relationships between the AM/FM signals, we apply the Isomap algorithm for unsupervised dimensionality reduction \cite{tenenbaum2000global}.
Isomap assembles a geodesic distance matrix by using neighborhood relationships from high-dimensional Euclidean distances. We first compute the MFCCs, Scattering1D and JTFS coefficients, STRFs and OpenL3 embeddings over the dataset of AM/FM signals.
Under the Isomap algorithm, we consider each representation separately.
To compute the nearest neighbor graph, we consider the 40 nearest neighbors for each transformed data point.
We select three components for the manifold visualization. The audio dataset described in Section \ref{sec:dataset} characterizes three independent degrees of freedom, therefore we postulate that Isomap will reveal whether the coordinates of an audio representation reflect similarities within the AM/FM signals. 
Musical instrument playing techniques are a class of signals that vary significantly in amplitude and frequency modulation content.
A recent publication showed that distances on the $K$-nearest-neighbor graph of JTFS reflected human similarity judgements between playing techniques \cite{lostanlen2021time}. 

We compute time--frequency scattering coefficients by means of our newly introduced implementation.
We transform each of the 4096 signals synthesized via Eq. (\ref{eqn:synth}), setting $J = 14$ octaves, $Q = 8$ filters per octave and $T = 1000~\rm{ms}$.
We omit frequential averaging to preserve equivariance to pitch transposition (see Eq.~(\ref{eq:jTFSTDefine-t})).
We set $J = 14$ to enable analysis of slower modulations via the temporal filterbank, with center frequencies reaching approximately 0.1 Hz.
We also compute time scattering (Scattering1D) coefficients using Kymatio \cite{andreux2020kymatio}, setting $Q = 1$ and $J = 14$ with global temporal averaging.
Time scattering does not capture spectrotemporal patterns beyond a log-frequency interval $1/Q_f$, where $Q_f$ is the quality factor (ratio of center frequency to bandwidth).
Hence, by setting $Q = 1$, which results in $Q_f = 2.5$, we guarantee that the scalogram contains at least one amplitude modulation cycle, given a modulation frequency of at least \SI{4}{\hertz} and a chirp rate of at most 4 octaves per second.

\begin{figure}[h!] 
    \centering
    (a)\includegraphics[width=.96\columnwidth]{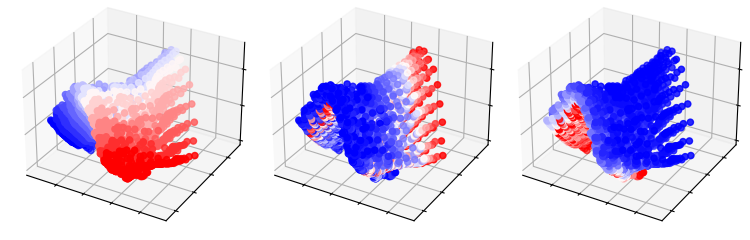}
    (b)\includegraphics[width=.96\columnwidth]{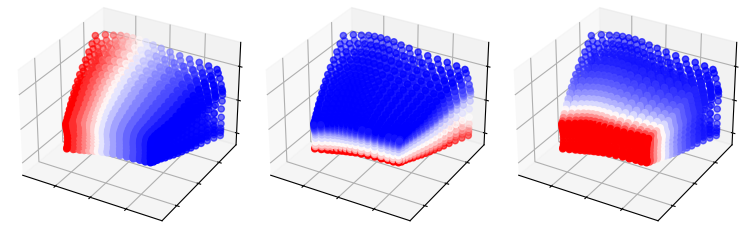}
    (c)\includegraphics[width=.96\columnwidth]{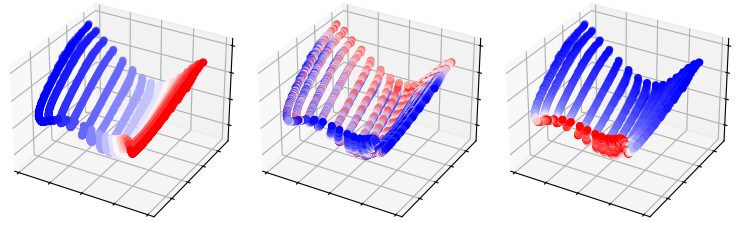}
    (d)\includegraphics[width=.96\columnwidth]{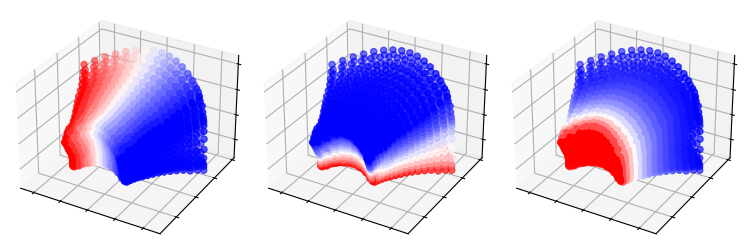}
    (e)\includegraphics[width=.96\columnwidth]{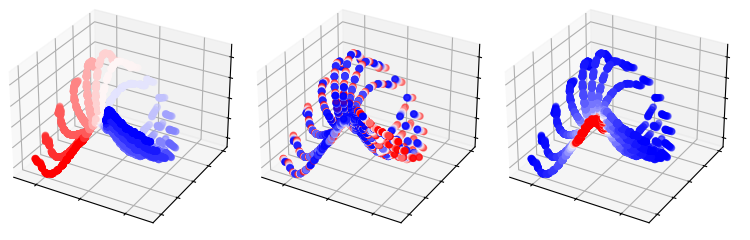}
    (f)\includegraphics[width=.96\columnwidth]{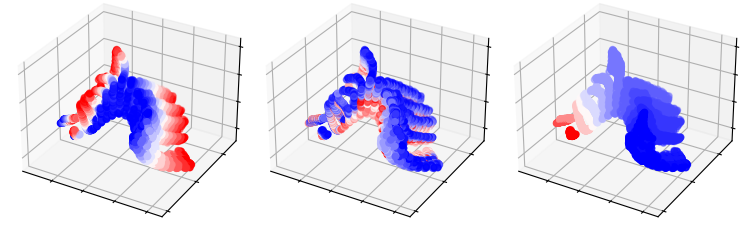}
    \caption{Isomap embeddings of a synthetic dataset of amplitude modulated sinusoidal chirps (see Section \ref{sec:dataset}) represented by: (a) MFCCs, (b) time scattering ($Q=1$) and (c) ($Q=8$), (d) time--frequency scattering, (e) STRFs and (f) OpenL3 embeddings. The configurations for these features are outlined in Section \ref{sec:isomap}. The embedding is fully unsupervised, using acoustic features alone. The colour range of the markers indicates an increasing value from blue to red via white, corresponding to the signals' carrier frequency $f_c$ (left), tremolo rate $f_m$ (center) and chirp rate $\gamma$ (right).}
    \label{fig:isomaps}
\end{figure}

Fig. \ref{fig:isomaps}(b) and (c) show three-dimensional (3-D) visualizations of the Isomap embeddings for time scattering ($Q=1$) and time--frequency scattering ($Q=8$), respectively. In the case of both transformations and the application of Isomap manifold learning, the dataset of AM/FM signals is represented as a 3-D mesh where the principal components align independently with $f_c$, $f_m$ and $\gamma$. Both transformations with their respective hyperparameters are capable of disentangling and linearizing fundamental frequency, tremolo rate and chirp rate, which describe spectrotemporal modulation patterns. Fig. \ref{fig:isomaps}(c) visualizes the embedding for time scattering when $Q = 8$. In this case, we observe that time scattering lies on a 2-D manifold that adequately describes $f_c$ and $\gamma$, yet fails to account for similarity in $f_m$ due to the aforementioned reasons. Despite time scattering successfully disentangling the 3 factors of variability when $Q = 1$ (Fig. \ref{fig:isomaps}(b)), other applications may demand a a greater quality factor in order to better localize in frequency.

As a comparison, we also compute Isomap embedding for the dataset's MFCCs (Fig. \ref{fig:isomaps}(a)), STRFs (Fig. \ref{fig:isomaps}(e)), and OpenL3 embeddings (Fig. \ref{fig:isomaps}(f)). We compute MFCCs using librosa V0.8 default parameters, yielding 20 coefficients \cite{mcfee2020librosa}. STRFs are computed by means of the `strf-toolkit` \cite{thoret2021learning} using the default parameters and setting the input duration to 4 seconds. OpenL3 embeddings are extracted using the \emph{music} model of the publicly available Python package\footnote{\url{https://openl3.readthedocs.io}}, resulting in 6144 coefficients after globally averaging in time.

We observe that in the case of MFCCs, the Isomap embedding forms a curved 2-D manifold, whereas our dataset contains three factors of variability. Only the fundamental frequency $f_c$ clearly aligns with one of the Cartesian coordinates. Meanwhile, similarities between amplitude modulation rates $f_{m}$ and chirp rates $\gamma$ are not represented faithfully. Therefore, neighboring points on the graph may have very dissimilar values of $f_m$ and $\gamma$. This is also the case for STRFs and OpenL3, where proximity relationships in the Isomap embeddings do not reflect similarity in tremolo rate $f_m$. In our experiment, the STRF fails to retrieve similarity in amplitude modulation rates. To determine the cause of this outcome demands a more thorough investigation as this behavior is contrary to its theoretical specification. Fundamental frequency and chirp rate are disentangled onto independent components, yet high chirp rates are densely clustered for all carrier frequencies.

\subsection{Regression from Nearest Neighbors}
As a quantitative supplement to the visualizations from the previous section, we assess regression of the synthesizer's three parameters $\boldsymbol{\theta}=(f_{c}, f_{m}, \gamma)$ with $K$-nearest neighbors regression algorithm ($K$-NN).
$K$-NN parameter regression relies on Euclidean distances between examples in their feature representations. Therefore, its regression error sheds light on the degree of topological alignment between feature space and parameter space. Such an alignment is essential in common audio recognition tasks, as the parameters are physical correspondents of audio similarity. 

\begin{figure}[h!]
    \centering
    \includegraphics[width=\columnwidth]{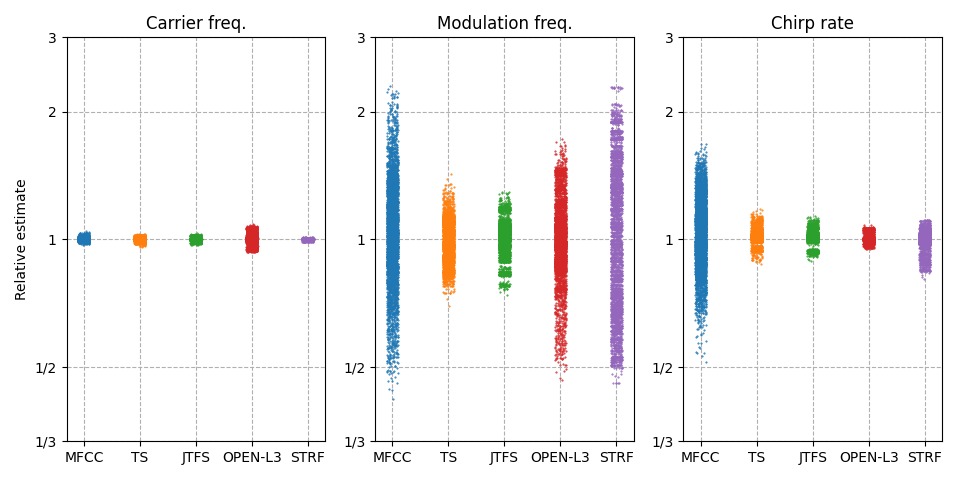}
    \caption{Error ratios that result from $K$-nearest neighbors ($K$-NN) regression of the AM/FM signal dataset's (Section \ref{sec:dataset}) three parameters: carrier frequency ($f_c$), tremolo modulation frequency ($f_m$) and chirp rate ($\gamma$). We performed K-NN regression via the nearest neighbor graphs ($K = 40$) that result from MFCCs, time scattering coefficients, time--frequency scattering coefficients, OpenL3 embeddings and STRFs. We refer the reader to Section \ref{fig:isomaps} for details on the feature extraction hyperparameters.}
    \label{fig:knn-regression}
\end{figure}

For each example, we start from an empty set of neighbors $\mathcal{N}_0 = \varnothing$. Then at each iteration, we select its closest neighbor by computing its pairwise Euclidean distance with all other examples. We stop after $K$ iterations, resulting in a set of $K$ nearest neighbors.
\begin{equation}
    \mathcal{N}_{k+1}(\boldsymbol{\theta}_{i}) =
    \mathcal{N}_{k}(\boldsymbol{\theta}_{i})
    \cup
    \left\{
    \argmin_{\boldsymbol{\theta}_{j} \not\in \mathcal{N}_{k}(\boldsymbol{\theta}_{i})}
    \left\Vert
    \mathbf{S}g(\boldsymbol{\theta}_{j})
    -
    \mathbf{S}g(\boldsymbol{\theta}_{i})
    \right\Vert_2
    \right\}
\end{equation}
We compute an estimate of the parameter $\widetilde{\boldsymbol{\theta}}_i$ as the average of its values at the $K$ nearest neighbors. We define the error ratio as $\widetilde{\boldsymbol{\theta}}_i/\boldsymbol{\theta}_i$, where:
\begin{equation}
    \boldsymbol{\widetilde{\theta}}_i =
    \dfrac{1}{K}
    \sum_{\theta_j\in\mathcal{N}_K(\theta_i)}
    \boldsymbol{\theta}_j.
\end{equation}
We use the same $K=40$ nearest neighbor graph computed by the Isomap algorithm in the previous section. We regress each example's parameters for each of the audio representations and plot their error ratios in Fig. \ref{fig:knn-regression}. 
All feature representations are capable of regressing carrier frequency $f_c$ with error ratios close to $1$. However, larger performance discrepancies can be observed in modulation frequency and chirp rate. Aligned with our observations in Section \ref{sec:isomap}, time scattering and JTFS excel at linearizing modulation frequency in the Euclidean space, with error ratios within range of $0.75$ to $1.5$. Meanwhile, all features except MFCCs extract chirp rate within error ratios between $0.75$ to $1.25$.  

\section{Time--frequency Scattering 2-D: Convolutional Classifier} \label{sec:convnet}
Recent publications have demonstrated time--frequency scattering as a state-of-the-art feature extractor for music classification tasks, including detection of the type of instruments played \cite{Anden2019joint} and detection of playing techniques \cite{wang2020playing} on solo performances. This is achieved by learning a shallow linear layer over time--frequency scattering coefficients. Time--frequency scattering is yet to be explored as a frontend feature extractor to a deep convolutional neural network (convnet) classifier. Exploiting the 3-D structure of time--frequency scattering $(\lambda_2 = (\alpha, \beta, \theta), \lambda, t)$, where the response of joint second-order wavelet filters across time and frequency compose the channels, may enhance contrast in spectrotemporal variations across the time--frequency image. In this section, we seek to compare audio representations as a frontend to a convnet in a task of supervised classification of musical instrument solos.

Eq. (\ref{eq:cnn_layer}) describes the output of the first layer of 2-D convolution between the time--frequency scattering image $\mathbf{S}\boldsymbol{x}$ around $(\lambda_2, \lambda, t)$ and kernel $\boldsymbol{w}$, where the multiindex variable $\lambda_2$ represents the tuple $(\alpha, \beta, \theta)$.

\begin{equation}
\label{eq:cnn_layer}
    Y(\lambda_3, \lambda, t)  \sum_{\lambda_2,\delta,\tau} \mathbf{S}\boldsymbol{x}(\lambda_2, \lambda + \delta - 1, t + \tau - 1)
    \boldsymbol{w}(\lambda_3, \delta, \tau)
\end{equation}

\subsection{Dataset}

We perform supervised classification of isolated musical instruments from the Medley-solos-DB dataset \cite{lostanlen2018medley}. The task of musical instrument recognition has previously been benchmarked with convolutional networks in \cite{lostanlen2016deep} and joint time--frequency scattering in \cite{Anden2019joint}. Every example in the dataset consists of a fixed $2^{16}$ discrete-time samples at a sampling rate of 44.1 kHz, corresponding to approximately 3 seconds of audio. Each clip includes the presence of one musical instrument from a highly imbalanced taxonomy of 8 classes: tenor saxophone, trumpet, flute, clarinet, female singer, distorted electric guitar, violin and piano, of 123, 149, 155, 251, 318, 404, 2040 and 2401 training samples, respectively. The training, validation and test subsets consist of a total of 5841, 3494 and 12236 samples, respectively.
% piano                        2401
% violin                       2040
% distorted electric guitar     404
% female singer                 318
% clarinet                      251
% flute                         155
% trumpet                       149
% tenor saxophone               123
\subsection{Convolutional Network Feature Design}
We parametrize time--frequency scattering such that it yields a 3-D output that is $44 \times 32$ along log-frequency and time. To achieve this, we set $J = 13$ octaves for the first and second order temporal wavelet filterbanks. We set $Q = 16$ filters per octave at first order and $Q_2 = 1$ at second order. We perform frequential averaging with the lowpass filter $\boldsymbol{\phi}_{F}$ over a quarter of an octave with $F = 4$. We set the support of the temporal lowpass filter $\boldsymbol{\phi}_{T}$ to $T = 2^{11}$, which is applied to an input of $N = 2^{16}$ discrete time samples. The frequential wavelet filterbank has its own set of parameters. $Q_{fr}$ and $J_{fr}$ control the number of wavelets per octave and  number of octaves and maximal scattering scale, respectively. We set $Q_{fr} = 1$ and $J_{fr} = 6$.

\begin{figure}[ht]
    \centering
    \includegraphics[width=\columnwidth]{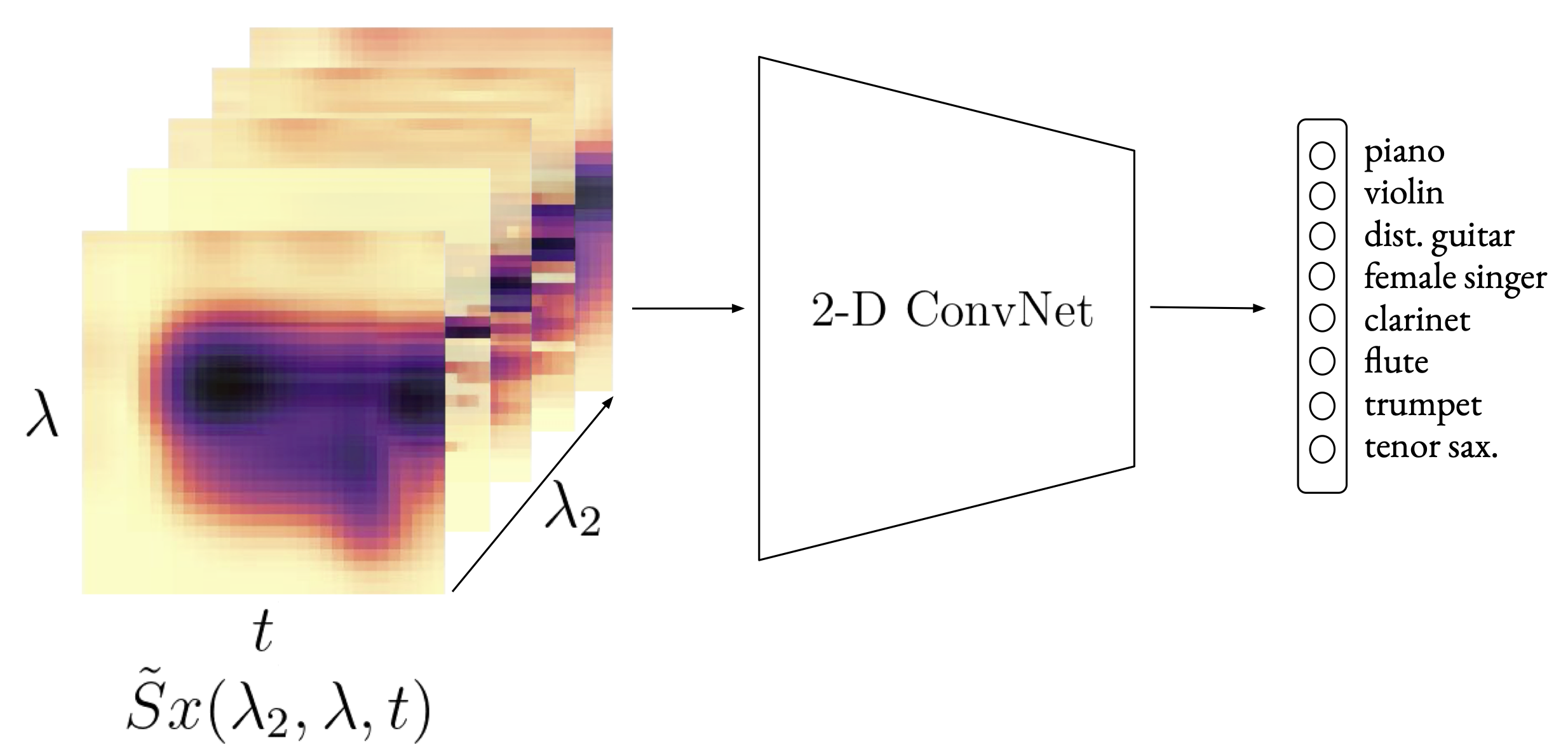} 
    \caption{2-D convnet architecture with a 3-D time--frequency scattering input tensor. The multiindex variable $\lambda_2 = (\alpha, \beta, \theta)$ represents the response of a second-order scattering wavelet at temporal rate $\alpha$, frequential scale $\beta$ and spin $\theta$. The 2-D convnet is an EfficientNetB0 architecture with a classification head over 8 classes (see Section \ref{sec:convnet} for details).}
    \label{fig:jtfs_conv}
\end{figure}

Our implementation supports outputs in various number of dimensions. In the case of 2-D output, the variables correspond to scattering path and time, and first-order and second-order coefficients are concatenated. In contrast, under the ``$\text{out\_3D}$'' mode, we return 2-D and 3-D tensors, for first-order and second-order tensors respectively.
Since $\boldsymbol{\phi}_{F}$ is not applied to first-order coefficients, we learn a convolution filter on the first order output and average pool across log-frequency.
We set the convolution filter's kernel size to $(16, 1)$, whose support covers an octave.
We apply average pooling to the convolution layer's output, with a kernel size of $(4, 1)$ corresponding to the same support of $\boldsymbol{\phi}_{F}$ of a quarter of an octave.
We concatenate the output of this layer with the 3-D second-order coefficients, resulting in a tensor of size $(195, 45, 32)$ whose indices correspond to log-frequency and log-quefrency unrolled, log-frequency and time.
We apply a 2-D batch normalization layer to this tensor, which serves as the input into a 2-D convnet.
Table \ref{tab:2dcnn} outlines our 2-D convnet that accepts the pre-processed time--frequency scattering tensor as input. We use a \emph{squeeze-and-excitation} layers \cite{hu2018squeeze} with a reduction factor of $16$.

\begin{table}[]
    \centering
    \begin{tabular}{c|c |c}
        \textbf{} & \textbf{channels} & \textbf{kernel size} \\ \hline
        Conv & $195$ & $(7, 7)$ \\
        Batch Norm, ReLU & $195$ &  \\
        Max Pool & & $(2, 2)$ \\ \hline
        Conv & $390$ & $(5, 5)$ \\
        Batch Norm & $390$ &  \\
        SqueezeExcite, ReLU & $390$ & \\
        \hline
        Conv & $780$ & $(3, 3)$ \\
        Batch Norm & $780$  &  \\
        SqueezeExcite, ReLU & $780$ & \\
        Global Average Pool & & \\ \hline
        Linear, ReLU, Dropout $0.5$ & $64$ & \\
        Linear, Softmax & $8$ &  \\
    \end{tabular}
    \caption{Table outlining the 2-D convnet architecture that process the time--frequency scattering tensor. The channels column indicates the number of channels or hidden units output after a layer.}
    \label{tab:2dcnn}
\end{table}

% \subsection{Adaptive Logarithmic Compression: AdaLog}
\subsection{Adaptive Logarithmic Compression} 

Prior to input to the convnet, we apply a transformation to each time--frequency scattering path, that seeks to match a decibel-like perception of loudness.
The transformation in Eq. (\ref{adalog}), $\mu$--$\log$, consists of mean-based renormalization and a pointwise logarithm.
We compute $\mu$ across the training set. When $\varepsilon$ is a non-learnable constant, we refer to this transformation as $\mu$--$\log$.
Previous publications have shown that for music sounds, $\mu$--$\log$ transforms each $\lambda_2$ such that its histogram of magnitudes is closer to Gaussian \cite{lostanlen2018extended}.
We set $\varepsilon$ to the same predefined constant $0.1$ per path, which was chosen based on our observations of the skewness of the magnitude histograms.
To standardize the input features to the convnet backend, we compute the mean and standard deviation per $\lambda_2$ across all $\widetilde{\mathbf{S}}\boldsymbol{x}$ in the training set.

\begin{equation}\label{adalog:mu}
    \mu(\lambda_2) = \frac{1}{N} \sum_{n=1}^N \iint \mathbf{S}\boldsymbol{x_n}(\lambda_2, \lambda, t) \;\mathrm{d}t\, \mathrm{d}\lambda
\end{equation}

\begin{equation}\label{adalog}
    \widetilde{\mathbf{S}}\boldsymbol{x}(\lambda_2, \lambda, t) = \log\Big(1 + \frac{Sx(\lambda_2, \lambda, t)}{\varepsilon\mu(\lambda_2)}\Big)
    % c(\lambda_2)
\end{equation}

\subsection{Baselines}
As a performance comparison to our time--frequency scattering hybrid convnet, we train a 2-D convnet on top of the Constant-$Q$ Transform (CQT) and a 1-D convnet on top of time scattering. 

We compute the CQT using nnAudio \cite{cheuk2020nnaudio} with a hop size of 256 samples, 96 frequency bins, 12 octaves and a minimum centre frequency of 32.7 Hz, and subsequently convert the amplitudes to the decibel scale. 
We standardize the CQT per bin using the means and standard deviations from the training set.
We perform average pooling over frequency and time with a kernel of size (3, 8), yielding a $(32, 32)$ time--frequency image for input into a 2-D convnet. To perform the classification, we use the same 2-D convnet that is outlined in Table \ref{tab:2dcnn}, however the successive convolution blocks have $64$, $128$ and $256$ channels respectively, the third convolution layer uses max pooling and all max pooling is performed with a kernel size of 2.

Additionally, we extract time scattering (Scattering1D) coefficients. Time scattering is computed similarly by applying only a 1-D temporal wavelet filterbank to the first-order scalogram. 
To match the setting of JTFS, we set $Q = 16$ filters per octave, $J = 13$ octaves and a temporal lowpass filter support of $T = 2^{11}$.
We concatenate first and second order coefficients, yielding a 1423-dimensional vector for each of the 32 time frames.
Note that we do not average the time scattering coefficients along first-order log-frequency $\lambda$.
We also apply the pathwise transformation $\mu$--$\log$ across time scattering paths. Time scattering is structured as a vector of scattering paths for each time frame $(p, t)$ where the path multiindex $p$ encompasses both scattering orders.
Hence, the integral in Eq. (\ref{adalog:mu}) is over the time variable alone. As the convnet classifier, we implement the same convnet as for CQT and JTFS, but with 1-D convolution, batch normalization and pooling operations.

% For 1--D scattering, we implemented a modified EfficientNet with 1--D convolutions. It accepts 1052 input channels. The first two layers consist of a 1--D convolution (512 and 256 kernels respectively) with a kernel size of 5, batch normalization, a pointwise ReLU nonlinearity, average pooling with a kernel size of two. An additional convolutional layer with 128 kernels follows, without average pooling. The convnet output is flattened and passed to a fully-connected layer of 84 units, dropout at a rate of $0.5$, then a linear classification layer with 8 logits.

\subsection{Training Setup}
We train the JTFS based models for 30 epochs and CQT and Scattering1D models for 20 epochs. We use an epoch size of 8192 and batch size of 32, by means of the AdamW optimizer with an initial learning rate of $10^{-3}$ and weight decay coefficient of $0.1$. To set the learning rate of each parameter in the network, we use a cosine annealing schedule with a minimum learning rate of $10^{-11}$, and apply warmup by starting at schedule's lowest point. In order to compensate for class imbalance in the Medley-solos-DB dataset, we use a weighted cross entropy (WCE) loss as per Eq. (\ref{eqn:wce}), where $N = \{N_1, ... , N_8\}$ is the set of training example supports per class.
\begin{equation} \label{eqn:wce}
    \text{WCE} = - \sum_i^B \frac{\max(N)}{N_{k = y_i} } y_i \log(f(x_i))
\end{equation}

\subsection{Results} \label{sec:results}
\begin{table*}[ht]
    \centering
    \begin{tabular}{|c| c c c c c c c c |c|}
    \hline
    & \textbf{tenor sax.} & \textbf{trumpet} & \textbf{flute} & \textbf{clarinet} & \textbf{female singer} & \textbf{dist. guitar} & \textbf{violin} & \textbf{piano} & \textbf{avg} \\ \hline
    \multirow{2}{*}{\textbf{CQT}}          & 5.7 & \textbf{80.9} & 40.2 & \textbf{85.3} & 84.4 & 87.8 & 64.2 & 98.5 & 68.4 \\
    & $\pm$ 3.8 & $\pm$ 3.5 & $\pm$ 5.3 & $\pm$ 5.1 & $\pm$ 0.7 & $\pm$ 1.8 & $\pm$ 12.8 & $\pm$ 1.4 & $\pm$ 1.8 \\\hline 
    \multirow{1}{*}{\textbf{Scattering1D}} & 54.8 & 70.9 & 43.9 & 60.1 & 93.7 & 96.1 & 74.1 & 98.7 & 74.0 \\ 
    & $\pm$ 9.8 & $\pm$ 9.7 & $\pm$ 8.9 & $\pm$ 7.9 & $\pm$ 2.4 & $\pm$ 1.1 & $\pm$ 4.3 & $\pm$ 0.2 & $\pm$ 2.4 \\ \hline
    \multirow{2}{*}{\textbf{JTFS}}         & \textbf{71.5} & 77.8 & \textbf{57.0} & 59.5 & \textbf{96.3} & \textbf{96.4} & \textbf{93.0} & \textbf{99.8} & \textbf{81.4} \\
    & $\pm$ 5.9 & $\pm$ 8.7 & $\pm$ 4.3 & $\pm$ 1.3 & $\pm$ 0.7 & $\pm$ 0.0 & $\pm$ 5.6 & $\pm$ 0.1 & $\pm$ 0.6 \\ \hline
    \end{tabular}
    \caption{Test set classwise and macro average accuracy for 2-D convnet architectures trained for musical instrument classification on Medley-solos-DB. Classes are in ascending order (left to right) of number of training set examples. We report the results of CQT, time scattering and time--frequency scattering frontends for a convnet classifier. See Section \ref{sec:convnet} for details.}
    \label{tab:cnn_acc}
\end{table*}

In Table \ref{tab:cnn_acc}, we report the classwise and macro-averaged accuracy on the test set. For each run, we use an early stopping procedure that checkpoints the model at each epoch. We select the best checkpoint out the final ten epochs that achieves the higher validation accuracy. We report the mean test set accuracy over three randomly seeded runs. 

Time--frequency scattering has previously seen performance of 78\% accuracy on the Medley-Solos-DB dataset when used with a shallow linear classifier \cite{Anden2019joint}. Earlier spiral convolutional network architectures achieved a highest average accuracy of 74\% \cite{lostanlen2016deep}. Our results show that the addition of a 2-D convnet backend exceeds the performance of a shallow linear classifier.  

JTFS exceeds the average accuracy of CQT and Scattering1D by roughly 23 and 7 percentage points respectively, while attaining the highest accuracy across the majority of musical instrument classes. We found that the unsupervised logarithmic transformation, $\mu$--$\log$, and careful selection of its tunable parameter $c$ were essential factors for improving performance. 

JTFS achieves state-of-the-art musical instrument classification performance in the regime of limited annotated data. As a reference, we compute accuracy metrics on the test set using a YAMNet classifier that was pretrained on the very large AudioSet dataset. This achieves 93\% accuracy with no additional trained layers. Yet we emphasise a distinction between these tasks; one is trained in the regime of limited annotated data, while the other has access to millions of annotated examples. The newly introduced implementation has enabled a previously unexplored interaction with learned 2-dimensional deep convolutional networks for supervised classification. We expect this to enable further research interfacing JTFS and convnets for audio analysis and synthesis.

\section{Texture Synthesis}\label{sec:textureresynth}
Under the context of an audio classification task, feature representations of waveforms benefit from invariance to time-shifting, pitch-shifting and small spectrotemporal deformations. Yet, the introduction of invariance induces inevitably a loss of information. Texture resynthesis from time-shift invariant feature is an effective way to examine what information is preserved and what is lost. Echoing work from  \cite{jtfs2015}, \cite{lostanlenflorian19}, and \cite{Anden2019joint}, the following section demonstrates the procedure of texture resynthesis, illustrates the texture preservation qualities of time-shift invariant JTFS, and reports the speed improvement of texture resynthesis with GPU-enabled JTFS-GPU over MATLAB toolbox \texttt{scattering.m}.

Starting from an audio signal $\boldsymbol{x}(t)$, we first calculate its scattering coefficients $\mathbf{S}\boldsymbol{x}$.
% apply scattering transform followed by a temporal low-pass filtering of time support $T$ to obtain its coefficients $\mathbf{S}\boldsymbol{x}$. 
To reconstruct the signal, we initialize a trial signal $\boldsymbol{y}(t)$ with random noise, and use backpropagation to update $\boldsymbol{y}$ at each iteration, such that the normalized error $E(\boldsymbol{y_n}) = \Vert\mathbf{S}\boldsymbol{x}-\mathbf{S}\boldsymbol{y_n}\Vert/\Vert\mathbf{S}\boldsymbol{x}\Vert$ is progressively reduced,
\begin{equation}
    \boldsymbol{y}_{n+1}(t) = \boldsymbol{y}_n (t) + \mu\boldsymbol{\nabla}E(\boldsymbol{y_n}).
\end{equation} 
While the loss function $E(\boldsymbol{y_n})$ is nonconvex and may have local minimizers, our goal is not to mathematically invert the forward scattering operations, but to approximate a sonification of scattering coefficients $\mathbf{S}\boldsymbol{x}$. 
The gradient $\boldsymbol{\nabla}E(\boldsymbol{y_n})$ is computed via reverse-ordered Hermitian adjoints of the forward scattering operations, detailed in \cite{lostanlenflorian19}.
We adopt the PyTorch backend for gradient computation, as well as a bold driver heuristic to update adaptively the learning rate $\mu$.
The bold driver heuristic increases the learning rate by a constant factor if the loss decreases and vice versa \cite{sutskever2013importance}. The reconstruction error decreases progressively: it reaches a normalized loss of about $10\%$ after 20 iterations and about $3\%$ after 100 iterations. 

\begin{figure}[h!]
    \centering
    \includegraphics[width=\columnwidth]{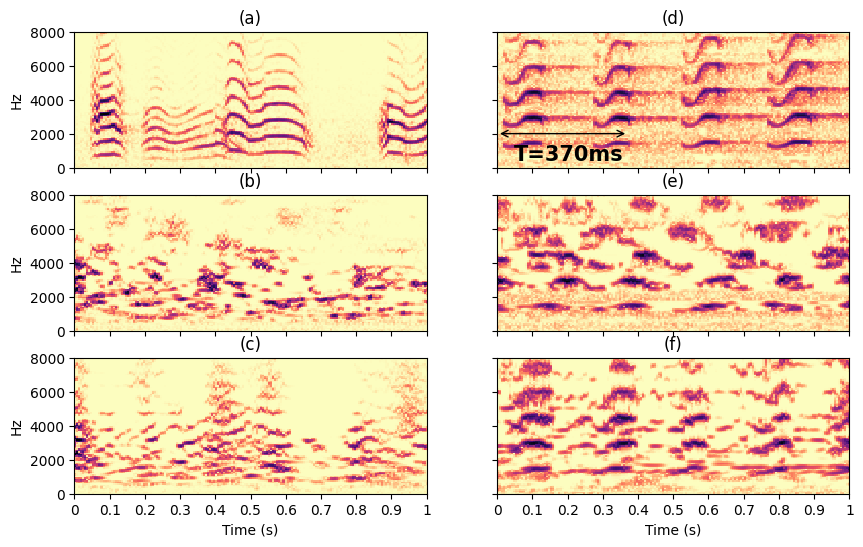}
   
    \caption{Reconstructed birdcalls from time-shift invariant scattering coefficients. (a) and (d) are scalograms of the original audio of two distinct birdcalls. (c) and (f) are the corresponding reconstructed scalograms from joint time--frequency scattering coefficients. (b) and (e) from second-order time scattering coefficients. All of the coefficients are computed with $J=12$, $Q=12$, enforcing time-shift invariance with a temporal lowpass filter of support $T=2^{13}$ samples (370 ms).}
    \label{fig:reconstruction}
\end{figure}

In order to illustrate the information that is preserved and lost in time-shift invariant JTFS, we experiment on the two bird chirps in Section \ref{sec:jtfs} (see Fig. \ref{fig:JTFS_example_visual} (c) and (d)) and compare their resynthesis results with those of second-order time scattering coefficients. The temporal support of time-shift invariance $T$ is chosen to be 370 ms, which is of the order of three bird calls in example (f) and thus relevant in a recognition task. We fix the scattering scale $J = 12$ and filters per octave $Q=12$ in all our experiments. As shown in Fig. \ref{fig:reconstruction}, both JTFS and time scattering preserve well the spacing along frequency axis and the amplitude modulation trend in each frequency bands. Meanwhile, both lose the precise temporal location of discrete bird call events because of temporal averaging. However, a clear distinction between them can be observed in terms of the frequency band alignment in time. Unlike time scattering, JTFS manages to recover synchronicity over frequency subbands.

To compare the computational speed, we perform texture resynthesis on an audio segment of size $N=2^{16}$ samples, i.e., around three seconds; both with JTFS-GPU and \texttt{scattering.m}. We record the time elapsed during each iteration of backpropagation. GPU computing accelerates the resynthesis procedure by close to ten times relative to \texttt{scattering.m}, with an average of 720 milliseconds per iteration.

\section{Conclusion}

Deriving auditory representations that act as proxies for perceptual similarity is an essential step for the enhancement of generative audio models and digital audio effects.
In this paper, we have highlighted the need for a scalable computational model of spectrotemporal receptive fields (STRF) of the auditory cortex.
We have provided \emph{scale--rate} visualizations of time--frequency scattering, analogous to those used in auditory perception research.
By means of practical examples, we have introduced a differentiable implementation of time-frequency scattering 
that is compatible with modern deep learning frameworks. Through manifold embedding visualizations and parameter recovery, we showed that time--frequency scattering can adequately serve as a representation of similarity for AM/FM signals.
By using our implementation's 3-D time--frequency scattering output as a frontend feature extractor for a 2-D convolutional neural networks classifier, we have exceeded previous state-of-the art benchmarks on the task of supervised classification of musical instrument solos with limited annotations.
Finally, we have demonstrated resynthesis of a variety of texture signals via their scattering coefficients, benefiting from a $10\times$ speedup over previous benchmarks.

\section{Acknowledgment}
We thank the DAFx 2022 organizing committee for their help.

\bibliographystyle{IEEEbib}
\bibliography{DAFx22_tmpl}
\end{document}